\documentclass[prl,twocolumn,aps,showpacs]{revtex4}

\usepackage{graphicx}

\begin{document}

\title{Atomic Fermi gas in the trimerised kagom\'{e} lattice
at the filling $2/3$}

\author{B. Damski$^{1,2}$, H.-U. Everts$^{1}$, A. Honecker$^{3}$,
H. Fehrmann$^{1}$, L. Santos$^{4}$ and
M. Lewenstein$^{1,5,*}$}

\affiliation{
(1) Institut f\"ur Theoretische Physik, Universit\"at Hannover,
Appelstr. 2, D-30167 Hannover\\
(2) Instytut Fizyki, Uniwersytet Jagiello\'nski, Reymonta 4,
PL-30-059 Krak\'ow\\
(3) Institut f\"ur Theoretische Physik, TU Braunschweig,
Mendelssohnstr. 3, D-38106 Braunschweig\\
(4) Institut f\"ur Theoretische Physik III,
Universit\"at Stuttgart, Pfaffenwaldring 57 V, D-70550 Stuttgart\\
(5) ICFO-Institut de Ci\`encies Fot\`oniques, Jordi Girona 29, Edifici Nexus II, E-08034 Barcelona, Spain
}
\begin{abstract}
We study low temperature properties of a spinless
interacting Fermi gas
in the trimerised kagom\'e lattice. The case of two fermions per trimer
is described by a quantum spin $1/2$
model  on the triangular lattice with couplings
depending on the bond directions.
Using exact diagonalisations  we show that the system exhibits
non-standard properties of a {\it quantum spin-liquid crystal},
combining a planar antiferromagnetic order with an exceptionally
large number of low energy excitations.
\end{abstract}

\maketitle

One of the most fascinating recent trends in the physics of
ultracold gases concerns
atomic gases in optical lattices,
where strongly correlated systems may be realized.
Such systems offer  an "atomic Hubbard toolbox" \cite{jaksch} to simulate
various sorts of  Hubbard models, and to study phenomena known in condensed
matter physics in an unprecedently controlled manner.
To name just few examples, atomic
lattice gases may serve to study various spin models \cite{demler},
to simulate
high $T_c$ superconductivity \cite{zoller},
to investigate a variety of quantum disordered
systems \cite{bosegl}, or to process quantum information \cite{phystod}.
Seminal experiments of Ref. \cite{bloch} have stimulated  a great interest
in experimental studies of atomic lattice gases (cf. \cite{bill}).

Particularly fascinating in this context is the possibility of studying
quantum frustrated antiferromagnets, which lie at the heart of modern
quantum magnetism \cite{review}. Recently we  have proposed how to create
ideal and trimerised kagom\'e optical lattices, and have studied physics of
various quantum gases in such lattices \cite{kagome}.
A Fermi-Fermi mixture with
half filling for both species in the limit of strong interspecies coupling
behaves in the ideal kagom\'e lattice as  a spin $1/2$ Heisenberg antiferromagnet (KAF).
Such a system (having so far no experimental realization among solid state systems)
is a paradigmatic, although not yet fully understood (c.f. \cite{nikolic})
example of a quantum  spin liquid of type II \cite{review}.

In Ref. \cite{kagome} we have also discussed
briefly
the case of interacting
spinless Fermi
gas in the trimerised kagom\'e lattice at  filling $2/3$
($2$ atoms per trimer).
Such a system behaves as a quantum magnet on the triangular
lattice with couplings that depend on bond directions, and is particularly interesting
since:
i)
it describes the physics of the trimerised KAF in the plateau region at one third of
the saturation magnetisation \cite{cabra04};
ii) it has itself fascinating properties, expected to be generic for
"multi"-merised frustrated systems;
iii) it is a paradigmatic
Fermi system to study in trimerised lattices;
iv) it is experimentally feasible.
In this Letter we
study low temperature $T$ physics  of this system  using exact
diagonalisations of the Hamiltonian for $12\cdots24$ spins.
We show that for effectively ferromagnetic couplings
the system exhibits
the
non-standard properties of a {\it quantum spin-liquid crystal},
 combining planar antiferromagnetic order with an
exceptionally large number
of low energy excitations, and a small (if any) gap.
For the effectively antiferromagnetic
coupling the {\em quantum} results  agree very well with the {\em classical} results
indicating  antiferromagnetic planar order and  a gapped spectrum.

The experimental  realization of the
considered system  requires a creation of trimerised
kagom\'e lattice,  using superlattice techniques as shown
in Ref. \cite{kagome}. The spinless interacting Fermi gas can then be
formed, for instance,  in a Bose-Fermi mixture, in the strong coupling
limit, when bosons form a Mott insulator (MI), and fermions
together with 0, 1, ... bosons
(bosonic holes) form fermionic composites \cite{fehrmann}.
Alternatively, one could use a gas of polarised ultracold
dipolar fermions, that interact via repulsive dipolar potential.

\begin{figure}[t]
\includegraphics[width=\columnwidth]{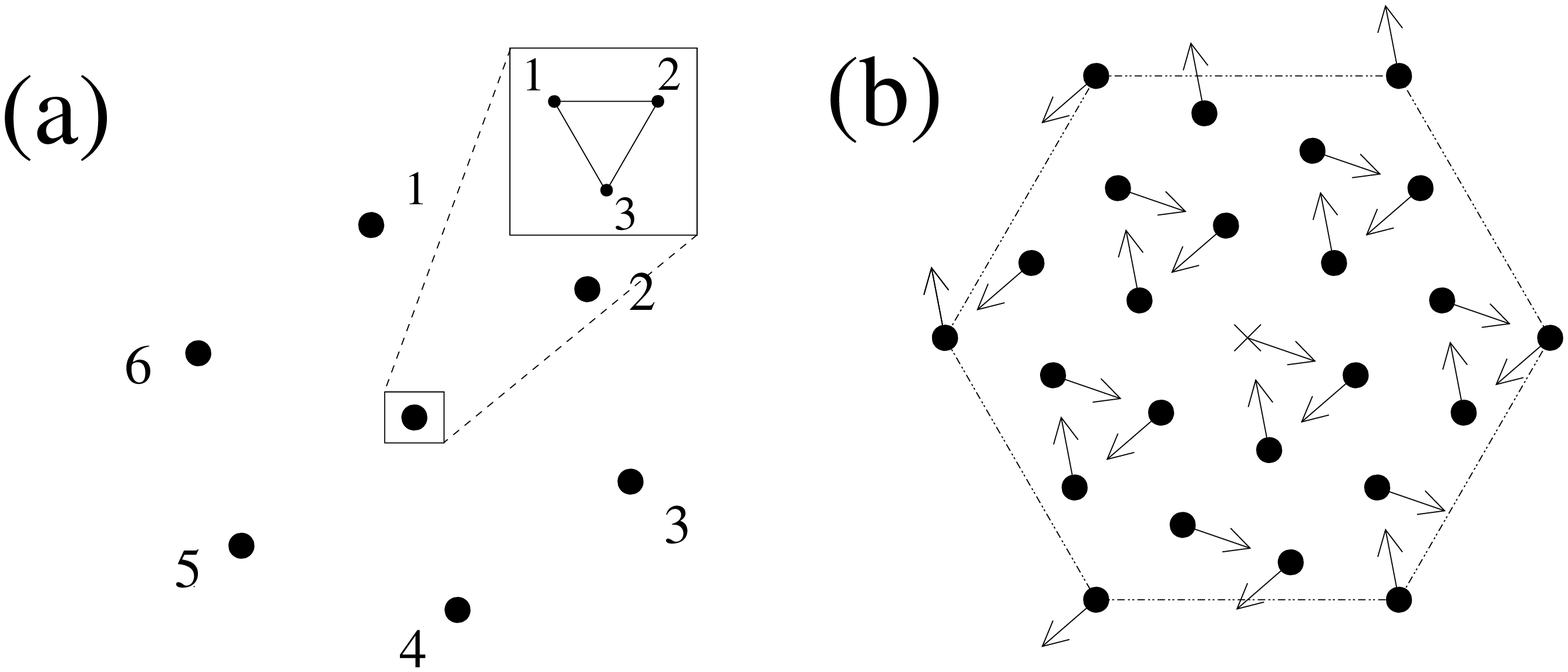}
\caption{(a) Enumeration of inter (intra) trimer nearest
neighbours; (b) Classical $120^\circ$ state  with {\it left} chirality.}
\label{fig1}
\end{figure}

The spinless interacting Fermi gas in the trimerised kagom\'e lattice
is described by the extended Fermi-Hubbard  Hamiltonian
$
H_{\mathrm{FH}}=-\sum_{\left\langle ab\right\rangle }
(t_{ab} f_a^{\dagger }f_b+ {\rm h.c.})
+\sum_{\left\langle ab\right\rangle}
U_{ab} n_{a}n_{b},
\label{FH}
$
where $a=\{\alpha,i\}$ with $\alpha$ referring to intra-trimer
indices and $i$ numbering the trimers.
The $t_{ab}$ and $U_{ab}$ take the values $t$ and $U$ for intra-,
and $t'$ and $U'$ for inter-trimer hopping,   $n_a=f_a^\dag f_a$,
and $f_a$ is the fermionic annihilation
operator.
The sites in each trimer
are enumerated as in Fig. \ref{fig1}a. We
denote the 3 different intra-trimer modes by
$f^{(i)}=(f_{1,i}+f_{2,i} +f_{3,i})/\sqrt{3}$
(zero momentum mode), and
$f_\pm^{(i)}=(f_{1,i}+z_{\pm}f_{2,i} +z_{\pm}^2f_{3,i})/\sqrt{3}$
(left and right chirality modes), where $z_{\pm}=\exp(\pm 2\pi i/3)$.

In the limit of weak coupling between the trimers of the original
kagom\'{e} lattice,
the problem of two fermions per trimer (filling $2/3$) becomes equivalent to a
quantum magnet on a triangular lattice with couplings
that depend on the bond directions as  described by the Hamiltonian
\begin{equation}
 H_{trimer} =
\frac{J}{2} \sum_{i=1}^{N} \sum_{j = 1}^6 s_i (\phi_{i \to j})
s_j (\tilde{\phi}_{j \to i}),
\label{subra}
\end{equation}
where $N$ denotes number of trimers,
$J=4U'/9$, and the nearest neighbours are enumerated as
in Fig.~\ref{fig1}a.
In  Eq. (\ref{subra}) we have
$s_i(\phi) = \cos(\phi) s_x^{(i)} + \sin(\phi) s_y^{(i)}$,
where the spin-$1/2$ operators
$s_x^{(i)}$, $s_y^{(i)}$ are defined as:
$s_x^{(i)} = (f_+^{(i)\dag} f_-^{(i)} +
f_-^{(i)\dag} f_+^{(i)})/2$,
$s_y^{(i)} = -i (f_+^{(i)\dag} f_-^{(i)}
-  f_-^{(i)\dag} f_+^{(i)})/2$. The angles $\phi$ are:
$\phi_{i \to 1} = \phi_{i \to 6} = 0$, $\phi_{i \to 2} = \phi_{i \to 3} = 2\pi/3$,
$\phi_{i \to 4} = \phi_{i \to 5} = -2\pi/3$,
$\tilde{\phi}_{i \to 1} = \tilde{\phi}_{i \to 2} = -2\pi/3$,
$\tilde{\phi}_{i \to 3} = \tilde{\phi}_{i \to 4} = 0$,
$\tilde{\phi}_{i \to 5} = \tilde{\phi}_{i \to 6} = 2\pi/3$.
This Hamiltonian has previously appeared  in the
context of a block-spin approach to the Heisenberg KAF
\cite{sub95, mila98}. The main purpose of that approach has
been to find the
origin of the  exponentially large number of low-lying singlets
that had been found in
numerical studies of the kagom\'{e} antiferromagnet \cite{lech97,waldt98}.
From Refs. \cite{sub95,mila98} it also follows that $H_{trimer}$ describes the physics
of the trimerised KAF in a magnetic field that drives this system into the plateau region at
$1/3$ of the saturation magnetisation.
We stress that the Hamiltonian ${ H_{ trimer}}$ to be studied in this Letter
 describes {\it a physically feasible  situation}.

\begin{figure}[t]
\includegraphics[width=\columnwidth]{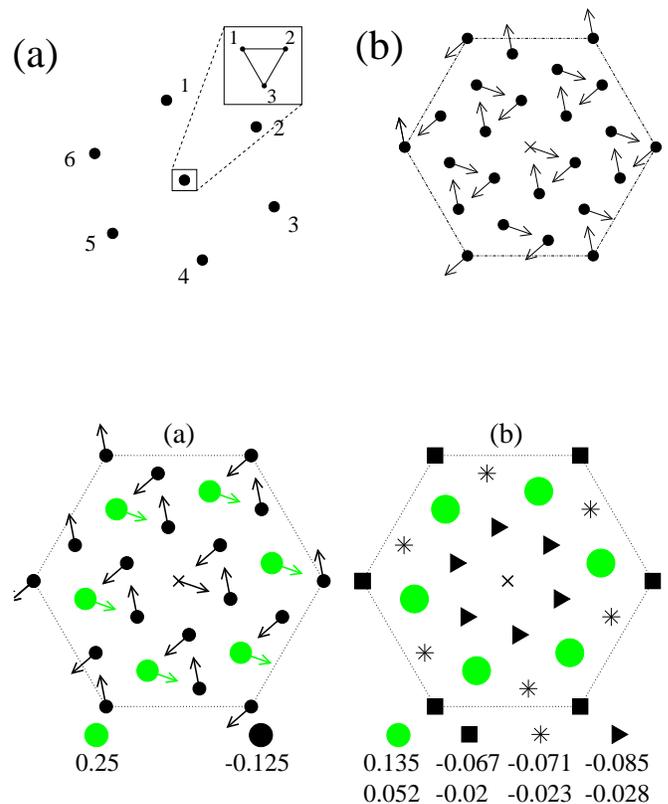}
\caption{(colour online).
(a) classical $120^\circ$ state with {\it right} chirality.
Dots show
$s_x^{(i)}s_x^{(10)}+s_y^{(i)}s_y^{(10)}$, where $|\vec{s}\;|=1/2$ \cite{x};
(b) spin-spin correlations,
$\langle s_x^{(i)}s_x^{(10)}+s_y^{(i)}s_y^{(10)}\rangle$.
The upper [lower] set of values corresponds to $kT=0$ [$kT=10^{-2}J/2$].
In both plots $N=21$ and  $i= 10$ at the central site.}
\label{fig2}
\end{figure}

Let us begin by discussing the classical theory of the model (\ref{subra}),
which describes the large spin limit.
In addition to being translationally invariant, the model of Eq.\ (\ref{subra}) is invariant
under the point
group of order 6, $Z_6=Z_3\cdot Z_2$, where the  generator of $Z_3$ (order 3)
is the combined rotation of the lattice by the angle $4\pi/3$, and of the spins
by the angle $2\pi/3$, while the generator of $Z_2$ (order 2) is the spin
inversion in the $x-y$ plane.
We remark that our model possesses no continuous spin rotational symmetry.
There exist three ordered classical states with small unit cells that
are compatible with this
point-group symmetry of the  model:
a ferromagnetic state, and two $120^{\circ}$ N{\'e}el type
structures with left (Fig. \ref{fig1}b) and right
(Fig. \ref{fig2}a) chiralities.
The energies
per site of these states are
 $E_{\rm class}^{\rm ferro} = E_{\rm  class}^{\rm  right} = -3 S^2 J/4$
 and $E_{\rm  class}^{\rm  left} = 3 S^2 J/2$,
where the subscripts ``right'' and
``left'' refer to  chiralities. Hence, for
$J<0$ the state with left-handed chirality will be the
ground state (GS). For $J>0$ the situation is more
complex: the states with
right-handed chirality and the ferromagnetic state are degenerate
ground states.

\begin{figure}[t]
\includegraphics[width=\columnwidth, clip=true]{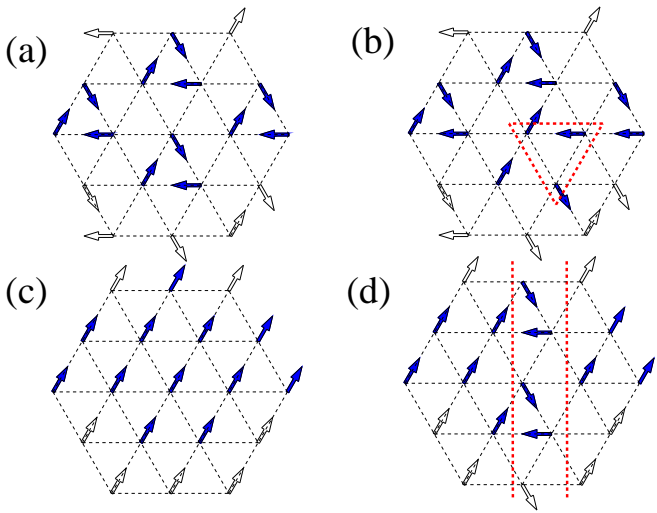}
\caption{(colour online)
(a)  Right chirality N\'{e}el configuration; (b) localised defect in configuration (a);
 (c) ferromagnetic configuration; (d) line  defect in
  configuration (c). Open arrows present spins
  determined by
  the
  boundary conditions for   the 12-spin cell.
Defects are marked by red dashed contours.
}
\label{defect}
\end{figure}

To understand further the nature of classical ground states
we have done a numerical analysis of the 12-spin cell by
fixing the direction of every spin to  $n\pi/3$ ($n=0\cdots5$),
and checking  the energies of the resulting
$6^{12}$ configurations. This analysis
has revealed that for $J<0$ there are $6$ ground states
($Z_6$ symmetry of (\ref{subra}))
each of them exhibiting the left chirality N\'{e}el order.
For $J>0$ the results are dramatically different:
there are $240$ degenerate classical GSs in this case
among them 6 pure right chirality  N\'{e}el states and 6 purely ferromagnetic states.
For an illustration,
we show in
Fig.~\ref{defect}
two ordered GSs with very large unit cells (Figs. \ref{defect}b, \ref{defect}d)
together with their parent states
(Figs. \ref{defect}a, \ref{defect}c).
As will be seen below
the large number of degenerate {\em classical} GSs finds its analogue in a
large density of low-lying excitations of the {\em quantum} version of  Eq. (\ref{subra}).

To describe the physics of spinless fermions on a trimerised optical
kagom\'{e} lattice
at  filling $2/3$ we need to consider the model,  Eq. (\ref{subra}),
for spin $1/2$,  {\it i.e.}
in the extreme quantum limit. Questions to be answered for this case are:
({\it i}) Is the GS of the model Eq.~(\ref{subra}) an ordered state,
or is it  a spin liquid
either of type I, {\it i.e.} a state
 without broken symmetry, with exponentially fast decaying
 spin-pair correlations
and a gap to the first excitation, or of type II, {\it i.e.} a
kagom\'{e}-like GS
again without broken symmetry, with extremely short ranged
correlations, but with a dense spectrum
of excitations adjacent to the GS;  ({\it ii})
What are the thermal properties of our
system? After all, the model can only be realized at
finite, albeit low temperatures.

To answer the above questions we have performed
exact diagonalisation of the Hamiltonian (\ref{subra})
for $N$= 12, 15, 18, 21, and 24 spins using the ARPACK
routines \cite{arpack}.
To simplify the calculations
we block-diagonalised the Hamiltonian (\ref{subra}) by exploiting
its translational symmetries thereby reducing the dimensions of the matrices
that had to be diagonalised from $2^N\times2^N$ to $\approx2^N/N\times2^N/N$.
Despite
all
these efforts,
studies of larger systems  require
the use
of massive
computer resources. Fortunately, the
results for 21 and 24 spins show qualitative and quantitative resemblance, and
we regard them as representative for larger systems.

\begin{figure}
\includegraphics[scale=0.31, clip=true]{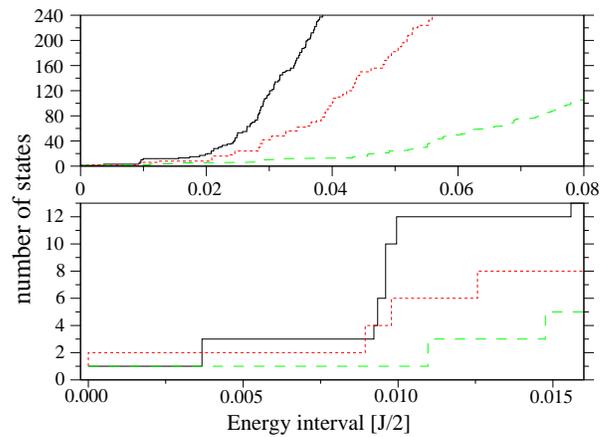}
\caption{(colour online). Number of states in an energy interval above
the GS: black (solid), red (dotted) and green (dashed) lines for $N=24$, $21$ and $18$, respectively.}
\label{fig3}
\end{figure}

Our findings are presented in Figs. \ref{fig2}b and \ref{fig3}
and Tables I and II.
For $J>0$,
 in contrast to the classical result,
the ground state exhibits the $120^{\circ}$
N\'eel order with right chirality \cite{chirality}. This
is illustrated in Fig. \ref{fig2}b, where
the planar spin-spin correlations
are presented.
Direct comparison with the correlations
of the classical state, Fig. \ref{fig2}a,
shows that  the exact quantum
correlations, although smaller, have the
same order of magnitude and sign as the classical ones. Especially,
the relative values of correlations compare nicely to the classical
result: Table I summarises the results for $N=21$ and $N=24$.
Amazingly, the $120^{\circ}$ N\'eel order survives at finite temperatures,
as is indicated by  the results obtained for $kT=10^{-2}J/2$, see Fig.~\ref{fig2}b.
At such temperatures about 800 low energy eigenstates contribute
to the  correlations. For smaller systems, $N<21$ ($J>0$), finite size effects
affect the spin
correlations
strongly. Nevertheless, the ground state energy per spin
can be reliably extracted from the data for  $N\ge12$, resulting in
$-0.2175J-0.0755 J/N$.

\begin{table}
\scriptsize
\begin{tabular}{|c|c|c||c|c||c|c||c|c||c|c|}
\hline
  &\multicolumn{2}{c||}{1} &
 \multicolumn{2}{c||}{$\sqrt{3}$} &
 \multicolumn{2}{c||}{2} &
 \multicolumn{2}{c||}{$\sqrt{7}$} &
 \multicolumn{2}{c|}{3} \\ \hline
$120^{\circ}$& -0.125 & 1.0  & 0.25 &2.0 & -0.125&1.0 & -0.125&1.0 & 0.25&2.0
\\ \hline
N=24 & -0.096& 1.0 & 0.162 & 1.69 & -0.083 &0.86 & -0.080 &0.83 &
0.156  &1.63\\ \hline
N=21 & -0.085& 1.0 & 0.135 &1.59 & -0.071 &0.84 & -0.067 & 0.79 & &
\\ \hline
\end{tabular}
\caption{Spin correlations for $J>0$.
For every distance from a reference centre ($1,\dots,3$  in lattice units)
the left most number is a planar spin-spin correlation, while
the right most one is the absolute value of that correlation divided
by the nearest neighbour spin correlation in the considered system.}
\end{table}

The selection of a GS with $120^{\circ}$ planar N\'{e}el order from the large
manifold of classical GSs by quantum effects implies the breaking of the translational
and of the point group of our model Eq.~(\ref{subra})
but there is no continuous
symmetry that the ordered GS  could break. Therefore,
the standard expectation would be that the excitations have a
gap of order of $J$. Instead, we find that
the system has an exceptionally large number of low energy excitations
(see Fig.~\ref{fig3}). For instance, for $N=21$ in the energy
interval $0.1J/2$ there are about $800$ excited states.
Most of them support the
spin order of the GS so that this order persists at finite temperatures.

The analysis  of the results for different $N$'s are compatible
with an  exponential increase of the number of
low-energy states with the system size $N$
similarly as in the case of the  $S=1/2$ KAF
\cite{lech97,waldt98}. For the KAF Mila  has
been able to explain this high density of low
energy states by associating  them approximately
to dimer coverings of an effective
triangular lattice with uncorrelated products of
nearest-neighbour pair states \cite{ mila98}.
His method fails here, because the low-lying states of our
model must certainly be
highly correlated.
On account of the breaking of the discrete symmetries of
our model
by the N\'eel order,  one expects the ground state of
the {\em infinite} system to be sixfold
degenerate.  For {\em finite} systems this degeneracy
is lifted. Nevertheless, we expect to find
six low-lying states in the gap below the lowest excited state.
In view of this scenario the inspection
of the lower panel of Fig. \ref{fig3}
suggests that the gap, if any,  is smaller than
$10^{-2}J/2$.
The appearance of this very small energy scale is completely
unexpected and puzzling. Obviously, the answer to the questions
({\it i}) and ({\it ii}) above, is that the GS is ordered, and that
the order survives at low $T$. The smallness of the gap and
the large density of low-energy states, however,
resemble very much the behaviour of a quantum spin liquid of type II.
For these reasons we propose to term our system a
{\it quantum  spin-liquid crystal}.
We remark in this context, that the specific heat of
$N=21$ system exhibits a peak at $kT \sim 7\cdot10^{-3}J/2$.

The above results for $J>0$ contrast dramatically with
those
for $J<0$, summarised in Table II. In the latter case we deal
with the standard quantum antiferromagnet with  $120^{\circ}$
N\'eel order and left chirality \cite{chirality} (Fig. \ref{fig1}b).
The spectrum is gapped, and the
classical spin-spin correlations approximate
the quantum correlations remarkably well.
The gap is of the order $|J|/2$ in this case ($N=12,18,21$), meaning that
there are at most a few states with energies substantially below $|J|/2$
for $J<0$, as opposed to the huge number for $J>0$ (Fig.\ \ref{fig3}).

\begin{table}
\scriptsize
\begin{tabular}{|c|c|c|c|c|}
\hline
  & $1$ & $\sqrt{3}$ & 2 & $\sqrt{7}$  \\ \hline
$120^{\circ}$& -0.125 & 0.25 & -0.125& -0.125 \\ \hline
N=21 & -0.134 & 0.237 & -0.117  & -0.116 \\ \hline
N=12 & -0.137 & 0.251 &  -0.125 & \\ \hline
\end{tabular}
\caption{Spin-spin planar correlations for $J<0$
as a function of distance $1,\dots,\sqrt{7}$ in lattice units.}
\end{table}

The observation of physics described in this
Letter requires achieving low, but not unrealistic
$T\simeq 10$nK to 100nK (c.f. Ref. \cite{fehrmann}). In experiments  $N$ could
vary from $\simeq 20$ to $\simeq 1000$.
The low energy states
may be prepared
by
 employing adiabatic changes of the
degree of trimerisation  of the lattice. For instance one can start with
a completely trimerised lattice; the filling $\nu=2/3$ may be achieved then
by starting with $\nu=1$, and eliminating 1 atom per trimer using, for
instance, laser excitations. One can then increase $t´$ and $U´$ slowly,
on the time scale slower than the final $1/J$ ($\simeq$seconds).
Alternatively, one  could start with $\nu\simeq 2/3$ in the moderately
trimerised regime. As in
Ref. \cite{bloch},  the inhomogeneity of the lattice due to the
trapping potential, would then allow to achieve the Mott state with
$\nu=2/3$ per trimer in the centre of the trap. Nearly perfect 2/3 filling can be achieved
by loading a BEC of molecules formed by 2 fermions into a triangular lattice, generating
an MI state,
adiabatically  transforming the lattice to a trimerised kagom\'e one, "dissociating" the molecules
by changing the scattering length to negative values,
and
by
finally optically pumping the atoms into a single internal state.
Preparing
$\nu=2/3$ might involve undesired heating (due to optical pumping),
which can be overcome by using laser, or phonon cooling afterwards (cf. \cite{daley}).
Note that the imperfections of $\nu$ can be described by a "$t-J$"-kind of model, and are
of interest themselves.

After state preparation it  should then be possible to measure
the energy of the system
simply by opening the lattice; by repeated
measurement of the energy $E(T)$ at (definite) finite temperatures one would
 get in this way an
access to the density of modes, i.e.  could compare the results with Fig.
\ref{fig3}. From such measurements one could infer
about the existence of the gap $E_{\rm gap}$, since if  $E_{\rm gap}$
 is large enough,  $E(T)$ becomes $T$-independent for $kT\le E_{\rm gap}$.
Various other
correlations could be measured using the
methods  proposed in Ref. \cite{ripoll}. In order to measure planar spin correlations,
one has, however, to lift
the degeneracy of the $f_{\pm}$ modes, e.g. by slightly modifying the
intensity of one of the superlattices forming the trimerised lattice.
This should be done on a time scale faster than the characteristic time
scales of other interactions, so that the state of the system would not
change during the measurement.  In such a case one can use far off
resonant Raman scattering (or scattering of matter waves) to measure the
dynamic structure factor, which is proportional to the spatio-temporal
Fourier transform of the density-density correlations. At frequencies
close to the two photon Raman resonance between the
$f_{\pm}$ modes, only $f_+$--$f_-$ transitions contribute to the
signal, and hence such measurement yields the desired information about
the correlations
$\langle f_+^{(i)\dag}f_-^{(i)}f^{(j)\dag}_-f_+^{(j)}\rangle$,
and the spin correlations of Fig. 2.

A.H. is indebted to D.C. Cabra and P. Pujol for collaboration in
a preliminary investigation  related to the present work. We acknowledge
support from the Deutsche
Forschungsgemeinschaft (SFB 407, SPP1116, 436 POL), ESF Programme QUDEDIS,
and the Alexander von Humboldt Foundation.

\end{document}